# Field-effect Modulation of Anomalous Hall Effect in Diluted Ferromagnetic Topological Insulator Epitaxial Films

Cui-Zu Chang,[1,2,3]* Min-Hao Liu,[2] Zuo-Cheng Zhang,[2] Yayu Wang,[2] Ke He,[1] and Qi-Kun Xue[2,1]

[1] State Key Laboratory of Low-Dimensional Quantum Physics, Department of Physics, Tsinghua University, Beijing 100084, China

[2] Beijing National Laboratory for Condensed Matter Physics, Institute of Physics, Chinese Academy of Sciences, Beijing 100190, China

[3] Francis Bitter Magnet Lab, Massachusetts Institute of Technology, Cambridge, MA 02139, USA

Email: czchang@mit.edu (C. Z. C)

**High quality chromium (Cr) doped three-dimensional topological insulator (TI) $Sb_2Te_3$ films are grown via molecular beam epitaxy on heat-treated insulating $SrTiO_3$(111) substrates. We report that the Dirac surface states are insensitive to Cr doping, and a perfect robust long-range ferromagnetic order is unveiled in epitaxial $Sb_{2-x}Cr_xTe_3$ films. The anomalous Hall effect is modulated by applying a bottom gate, contrary to the ferromagnetism in conventional diluted magnetic semiconductors (DMSs), here the coercivity field is not significantly changed with decreasing carrier density. Carrier-independent ferromagnetism heralds $Sb_{2-x}Cr_xTe_3$ films as the base candidate TI material to realize the quantum anomalous Hall (QAH) effect. These results also indicate the potential of controlling anomalous Hall voltage in future TI-based magneto-electronics and spintronics.**

Topological insulators (TIs) are a recently discovered class of quantum coherent materials with an insulating bulk but metallic surfaces/edges. TI's metallic Dirac surface states (DSSs) show many fascinating properties including: an extraordinary spin texture, massless Dirac fermions, and robust stability.[1-3] The tetradymite semiconductor diantimony tritelluride ($Sb_2Te_3$) has been theoretically predicted, and experimentally observed, to be a three-dimensional (3D) TI with single Dirac cone [4-6] that lies well inside the bulk gap. $Sb_2Te_3$ is always *p*-type due to the formation of native antisite defects.[5-7] $Sb_2Te_3$ also belongs to the family of narrow-bandgap semiconductors, with a bulk gap of ~260 meV. Doping transition metal elements in conventional narrow-bandgap semiconductors usually induces ferromagnetism. These doping-induced ferromagnetic systems are known as diluted magnetic semiconductors (DMSs).[22] We doped the matrix of $Sb_2Te_3$ with varying amounts of chromium (Cr), which can lead to a robust long-range ferromagnetic state.[8] It is well known that, time-reversal symmetry breaking in 3D TI-based tetradymite-type DMSs yields novel electronic and magnetic properties including: gap opening at the Dirac point (DP);[9] the quantum anomalous Hall (QAH) effect;[10-12] the topological magneto-electric (TME) effect;[11] and image magnetic monopoles.[13]

The ultrathin Cr-doped $Sb_2Te_3$ films were grown by molecular beam epitaxy (MBE) in an ultra-high vacuum (UHV) with a base pressure better than $2 \times 10^{-10}$ mbar.[14] Insulating commercial $SrTiO_3$(111) substrates were used because their large dielectric constant after heat treating ($\kappa$ ~30000 at temperature $T$~2K), enables tuning carrier density ($n_{2D}$) with a bottom gate bias ($V_g$), even with a thick (~0.5mm) substrate.[15] High purity Sb (99.9999%), Te (99.9999%) and Cr (99.999%) were simultaneously evaporated from standard Knudsen cells. Unlike to the MBE growth of $Bi_2Se_3$ or $Bi_2Te_3$ films, [14,16-17] we can obtain high quality stoichiometric Cr doped $Sb_2Te_3$ films with a (Sb, Cr)/Te flux ratio is only set about 1:3~1:2 and a substrate temperature set at 180°C. The typical growth rate in our experiment is ~ 0.33 QL/min, when the Sb and Te sources are set to 380°C and 260°C, respectively. **Fig.1a** shows the reflective high-energy electron diffraction (RHEED) patterns of the heat-treated

SrTiO$_3$(111) substrate for a $[2\bar{1}\bar{1}0]$ incident beam. The sharp reconstruction and clear Kikuchi lines demonstrate the high crystalline quality of the surface. **Fig. 1b** shows the RHEED patterns of 5 QLs Sb$_{1.91}$Cr$_{0.09}$Te$_3$ films on the SrTiO$_3$(111). A clear and sharp 1×1 pattern appears, and sharp streak-like RHEED patterns attest to a two dimensional (2D) layer-by-layer mode of thin film growth.[14] In Cr doped Sb$_2$Te$_3$ films, the Cr atom neither substitutes on the anion site of Te nor enters inside the van der Waal gap, it only substitutes on the Sb site.[8, 18]

To avoid possible ambient contamination of TI films, a 20nm-thick insulating amorphous Te capping layer was deposited on top at low temperature $T$~150K prior to taking the films out of the UHV chamber. Both the Te capping layers and the substrates are excellent electrical insulators, so their contributions to electrical transport can be safely neglected in electrical transport measurements. After removing the films from the chamber, 10nm Ti and 100nm Au electrodes were deposited through a shadow mask on the top of the sample to make ohmic contact, and to form Hall bar geometry. Low temperature silver conductive adhesive was overlaid on the bottom of the SrTiO$_3$(111) as a bottom gate electrode. The final millimeter-sized transistor devices are schematically shown in **Fig.1c**.[19] Robust long-range ferromagnetic states in our Cr doped Sb$_2$Te$_3$ films were demonstrated through both electrical transport measurements (**Fig.1d**) and direct magnetization measurements with a superconductivity quantum interference device (SQUID) magnetometer (**Fig.1e**). **Fig.1d** shows $T$ dependent Hall traces of 10QL Sb$_{1.7}$Cr$_{0.3}$Te$_3$ films. Clear square hysteresis loops at low temperatures ($T$<55K) show that the 10QL Sb$_{1.7}$Cr$_{0.3}$Te$_3$ films are in a ferromagnetic state. The inset of **Fig.1d** is a longitudinal resistance ($R_{xx}$) vs $T$ curve, which shows a local maximum (hump) around 55K. Such behavior in the $R_{xx}$-$T$ curve is due to spin disorder scattering that sets in at the paramagnetism-to-ferromagnetism transition. This scattering indicates the magnetic ordering temperature, known as the Curie temperature ($T_C$).[9] $T_C$ of 10QL Sb$_{1.7}$Cr$_{0.3}$Te$_3$ films is ~55K. **Fig.1e** shows the magnetization-field curves of 100QL Cr$_{0.22}$Sb$_{1.78}$Te$_3$ films. These curves show perfect hysteresis loops, where the magnetic moment per Cr ion is determined by the saturation magnetization to be ~3.4$\mu_B$. This indicates that Cr atoms in the films form $Cr^{3+}$ ions. The inset of **Fig.1e shows** the temperature dependent remanent magnetization exhibits an abrupt upturn around $T \sim 40K$. This indicates the $T_C$ of 100QL $Cr_{0.22}Sb_{1.78}Te_3$ films is ~40K. $T_C$ can be increased to high temperature with increasing Cr content in matrix of $Sb_2Te_3$.

The evolution of electronic band structure resulting from different Cr concentrations in 5QL $Sb_{2-x}Cr_xTe_3$ films on $SrTiO_3$ (111) has been revealed by *in situ* angle-resolved photoemission spectroscopy (ARPES) measurements. All ARPES band maps were taken along the $\bar{K}$-$\bar{\Gamma}$-$\bar{K}$ direction at $T$~150K. To avoid sample charging during ARPES measurements, a 300-nm-thick titanium (Ti) film is deposited at both ends of the substrate. Once a continuous film is formed, the sample is grounded to the sample holder through the contacts.[14] **Fig.2a** shows the ARPES band spectrum of pure $Sb_2Te_3$, which has well defined DSSs and a DP that lies at 65meV above the $E_F$.[18] Different doping levels ($x$~0.05, 0.09 and 0.14) were measured, and **Figs.2b** to **2d** show that the Cr doping causes insignificant changes to the DSSs. The topological DSSs persist with increasing $x$; meanwhile, the energy difference between the DP and $E_F$ becomes moderately larger with increasing $x$. This result demonstrates that a small number of hole-type carriers are introduced by Cr doping.

Electrical transport measurements have been performed in a cryostat with magnetic field ($\mu_0H$) up to 15T and $T$ down to 1.5 K, respectively. The Hall resistance ($R_{yx}$) and $R_{xx}$ were measured using a standard ac lock-in method with the current parallel to the film plane and the magnetic field applied perpendicular to the film. A series of samples with progressively increasing content of Cr doping and electric field gating strengths were measured. Hall traces show that all $Sb_{2-x}Cr_xTe_3$ films have hole-type carriers and display a similar gate field-effect on anomalous Hall resistance. Because of these similarities, we take $x= 0.09$ as an example system, and all further data presented here is recorded from this sample.

**Fig.3a** shows $\mu_0H$ dependence of the $R_{yx}$ at $T$=1.5K in the series of gate voltages $V_g$ = -210V, -100V, -50V, 0V, +50V, +100V, and +210V. Clear hysteresis loops show that the 5QL $Sb_{1.91}Cr_{0.09}Te_3$ films are in a robust long-range ferromagnetic state. The hysteresis loops also indicate that the easy magnetization axis is perpendicular to the film, which is more useful for spintronic devices applications compared with an in-plane easy magnetization axis.[8] The vertical intercept of the hysteresis loop becomes lager when $V_g$ changes from -210V to +210V; the vertical intercept at $V_g$=+210V is about twice of that at $V_g$=-210V. The inset of data in **Fig.3a** shows the $R_{xx}$ vs $V_g$ at $T$=1.5K, and shows that $R_{xx}$ increases as $V_g$ increases. Assuming a constant mobility, we can conclude that the total $n_{2D}$ becomes lower with $V_g$ from -210V to +210V. In a 2D system, the total $R_{yx}$ can be defined as $R_{yx}=R_A \cdot M + R_O \cdot \mu_0 H$. Here $R_A$ is the anomalous Hall coefficient, $R_O$ is the ordinary Hall coefficient, $M$ is the magnetization perpendicular to the film. In the low magnetic field limit, the anomalous Hall effect component dominates the Hall resistance; whereas in the high magnetic field limit the ordinary Hall effect component dominates the Hall resistance. The anomalous Hall resistance ($R_{AH}$) is determined by the extrapolation of the component linear in magnetic field. The slope of $R_{yx}$ vs $\mu_0 H$ curve at low $T$ under high magnetic field, where magnetization saturates, determines the type of conducting carriers and the total $n_{2D}$. In our samples, this deviation from the linear behavior occurs below 1T, which coincides with the saturation field of magnetization. $R_{AH}$ and the total $n_{2D}$ as a function of $V_g$ at $T$=1.5K are displayed in **Fig.3b**. Both $R_{AH}$ and $n_{2D}$ show a $V_g$ dependence. $R_{AH}$ changes from 24.1 Ω to 40.8 Ω, and $n_{2D}$ changes from $3.4 \times 10^{14} cm^{-2}$ to $9.7 \times 10^{13} cm^{-2}$ as $V_g$ increases from -210V to +210V. Intriguingly, $R_{AH}$ increases as $n_{2D}$ decreases. This phenomena, which results from a giant van Vleck magnetic susceptibility of valence electrons in TI systems,[10] is different from the machanism of ferromagnetism in conventional DMSs (*e.g.* (Ga,Mn)As) in which the carrier mediated and Ruderman-Kittel-Kasuya-Yosida (RKKY) interactions are responsible for the appearance of ferromagnetism and lead to a positive correlation between $R_{AH}$ and $n_{2D}$.[20-21]

**Figs.4a** to **c** display the variation of $R_{yx}$ vs $\mu_0 H$ with gate voltages $V_g$= -100V, 0V, +210V, and $T$ varied from 1.5K to 50K. Clear hysteresis loops can be observed in the region 1.5K≤$T$<15K. As $T$ increases the hysteresis loops vanish and the Hall traces display nonlinear behavior at low field (15K≤$T$<25K). At $T$≥25K the nonlinearity vanishes and the Hall traces become linear. The insets in **Figs. 4a** to **c** show the $R_{xx}$ vs $T$ curves for $V_g$= -100V, 0V, +210V. The film is metallic at $V_g$= -100V, but it displays insulating behavior at $V_g$ = +210V. All three $R$-$T$ curves show an increase in $R$ at low temperature due to gap opening induced by ferromagnetism at low temperature.[9]

The $T$ dependence of the remnant resistance ($R_H$) in $R_{yx}$ at $\mu_0H$ =0T at $V_g$ = -100, 0, +210V is shown in **Fig.5a**. There is little change in $T_C$ as $V_g$ is varied at $T$=12.5K and $T$=15K. The inset of **Fig.5a** displays a typical Arrot plot of the film, at $V_g$ =0V. The Arrot plot accurately confirms the occurrence of ferromagnetic state, since the effect of magnetic anisotropy and domain rotation can be minimized. The intercept of the high field isotherms of $R_{yx}^2$ vs $H/R_{yx}$ give the spontaneous $R_{yx}$ at $\mu_0H$ =0. The isotherm at $T_C$ passes though the origin (0, 0) point, and a negative intercept of an isotherm means no magnetiztion. The $T_C$s of the film at these three $V_g$s are all close to 15K, as the intercept of the linear component of the fits to the $R_{yx}^2$ vs $H/R_{yx}$ curves around 15K is very close to the origin.[23] The temperature dependence of coercivity field strength ($H_c$) of the film at $V_g$ = -100, 0, +210V is shown in **Fig.5b**. The values of $H_c$, and therefore also the saturation magnetizations, are almost identical at constant $T$ with variation in $V_g$. The saturation magnetizations at three $V_g$s are near 130mT along the easy magnetization axis. **Fig.5c** shows the 2D ordinary Hall coefficient, $R_O^{2D}$, as a function of $T$. $R_O^{2D}$ can be expressed as $R_O^{2D} = \dfrac{1}{n_{2D}H}$. The $T$ dependence of $n_{2D}$ is obtained from this expression, and we see that $n_{2D}$ initially decreases, and then increases with the decreasing temperature below $T$=20K. This $T$ dependence is the typical behavior of conventional semiconductors.

To summarize, an insensitive DSS and a robust long-range ferromagnetic state have been unveiled in ultrathin epitaxial films of Cr doped $Sb_2Te_3$. These characteristics were confirmed by *in situ* ARPES measurements and the presence of hysteresis loops in the anomalous Hall effect and SQUID magnetometer, respectively. We can tune the anomalous Hall effect through the effective bottom field-effect gate. While $R_{yx}$ increases with the decreasing $n_{2D}$, $H_c$ is not significantly changed. This indicates a robust long-range ferromagnetic state that is insensitive to the carrier type and density, which is different from the ferromagnetic order mechanism in conventional DMSs. Because of this weak, carrier independent ferromagnetism in Cr doped $Sb_2Te_3$, we chose Cr doped $Sb_2Te_3$ as the base candidate TI material to successfully realize the QAH effect.[12, 18] This work also paves an effective path for controlling anomalous Hall voltage of magnetically doped TI thin films in future dissipationless TI-based magneto-electronic and spintronic devices.

## Acknowledgment

We thank R. Wu for fruitful discussions. This work was supported by the National Natural Science Foundation of China, the ministry of Science and Technology of China, and the Chinese Academy of Sciences.

## References

1. M. Z. Hasan and C. L. Kane, Rev. Mod. Phys. 82, 3045-3067 (2010).
2. X. -L. Qi and S. -C. Zhang, Rev. Mod. Phys. 83, 1057 (2011).
3. J. E Moore. Nature 464, 194-198 (2010).
4. H.J. Zhang, C. X. Liu, X. L. Qi et al. Nature Phys. 5, 438-442 (2009).
5. D. Hsieh et al. Phys. Rev. Lett. 103, 146401 (2009).
6. G. Wang et al. Nano Res. 12, 874-880(2010).
7. Y. P. Jiang et al. Phys. Rev. Lett. 108, 066809 (2012).
8. Y. J. Chien, thesis, University of Michigan, Ann Arbor, MI (2007).
9. Y.L. Chen et al. Science 329, 659-662 (2010).
10. R. Yu et al. Science 329, 61-64 (2010).
11. X. L. Qi et al. Phys. Rev. B 78, 195424 (2008).
12. C. Z. Chang et al. Science 340, 167 (2013).
13. X. L. Qi et al. Science 323, 1184(2009).
14. C. Z. Chang et al. SPIN 1, 21 (2011).
15. J. Chen et al. Phys. Rev. Lett. 105, 176602 (2010).
16. Y. Zhang et al. Nat. Phys. **6**, 584 (2010).
17. Y. Y. Li et al. Adv. Mater. 22, 4002 (2010).

18. C. Z. Chang et al. Adv. Mater. 25, 1065 (2013).
19. M. H. Liu et al. Phys. Rev. B 83, 165440 (2011).
20. H. Ohno et al. Nature 408, 944-946(2000).
21. D. Chiba et al. Appl. Phys. Lett. 89, 162505(2006).
22. H. Ohno. Science 281, 951 (1998).
23. A. Arrott. Phys. Rev. 108, 1394 (1957).

**Figures and figure captions**

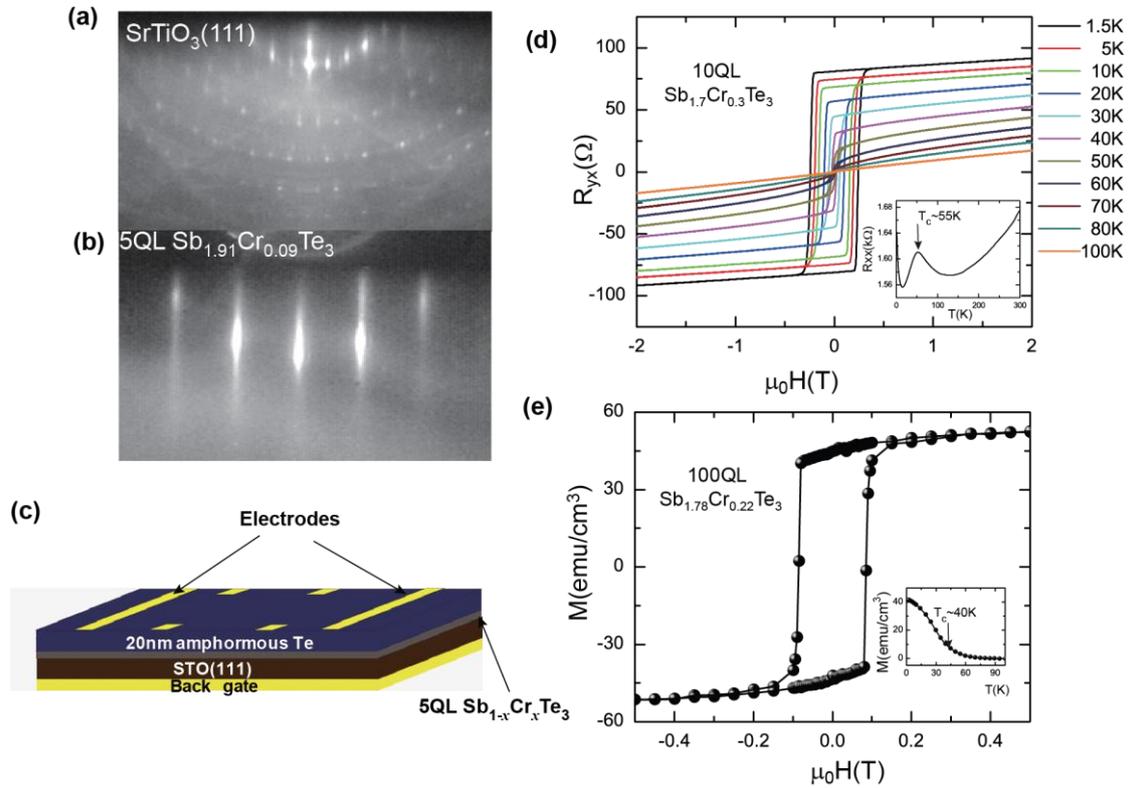

**Fig.1** RHEED patterns of **(a)** heat-treated SrTiO$_3$(111) and **(b)** 5QL Sb$_{1.91}$Cr$_{0.09}$Te$_3$ films. **(c)** Schematic of the 5QL Cr-doped Sb$_2$Te$_3$ films for the transport measurements (the thickness is not to scale). **(d)** Magnetic field ($\mu_0H$) dependent Hall resistance ($R_{yx}$) of the 10 QL Sb$_{1.7}$Cr$_{0.3}$Te$_3$ films at varied temperature, inset, the longitudinal resistance ($R_{xx}$) vs $T$ curve. **(e)** SQUID hysteresis loops of 100QL Sb$_{1.78}$Cr$_{0.22}$Te$_3$ at $T$=2K, inset, $T$ dependent remanent magnetization ($M$) curve.

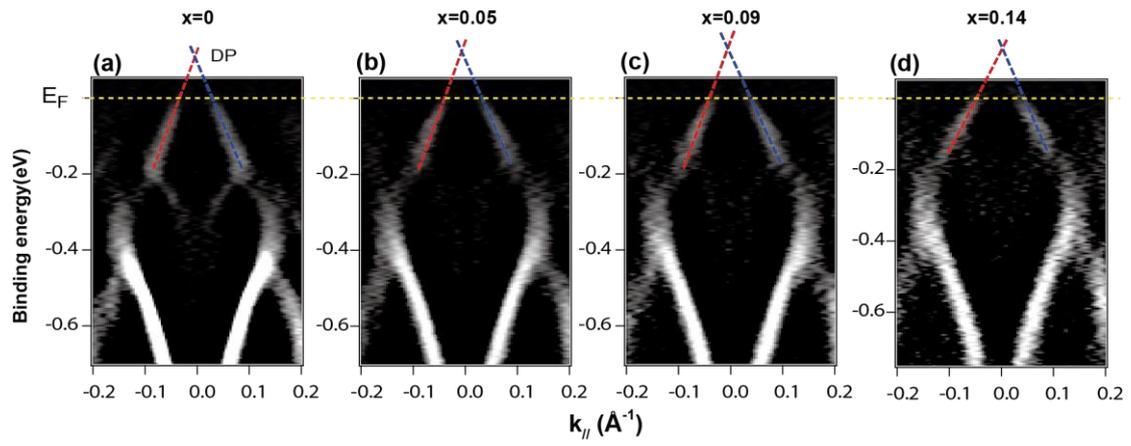

**Fig. 2** ARPES band spectra of $Sb_{2-x}Cr_xTe_3$ taken along the $\bar{K}$-$\bar{\Gamma}$-$\bar{K}$ direction at $T\sim150K$ for: **(a)** $x$=0; **(b)** $x$ =0.05; **(c)** $x$ =0.09; and **(d)** $x$ =0.14.

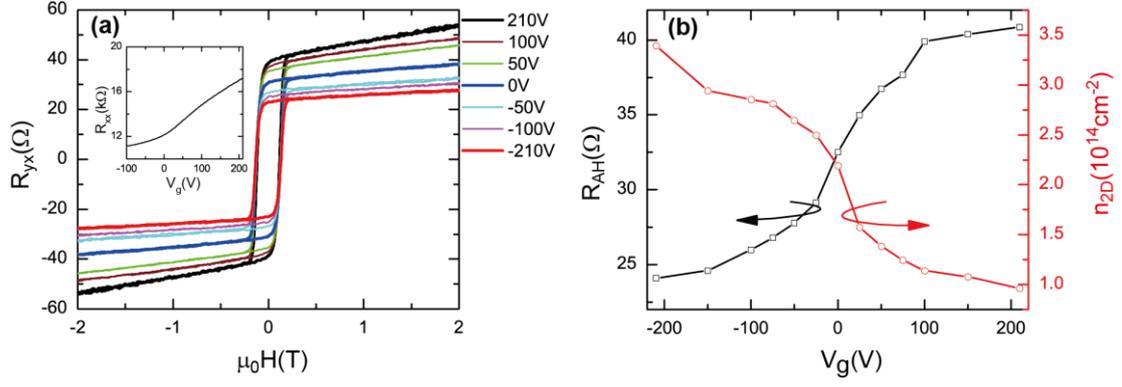

**Fig.3** $\mu_0H$ dependence of the Hall traces of 5QL $Sb_{1.91}Cr_{0.09}Te_3$ at $T=1.5K$ with different gate bias ($V_g$). **(a)** $R_{yx}$ vs $\mu_0H$ curves for $V_g$ from -210V to +210V. Inset, $R_{xx}$ vs $V_g$ curve at $T=1.5K$. **(b)** The anomalous Hall resistance ($R_{AH}$) and the carrier density ($n_{2D}$) as a function of $V_g$.

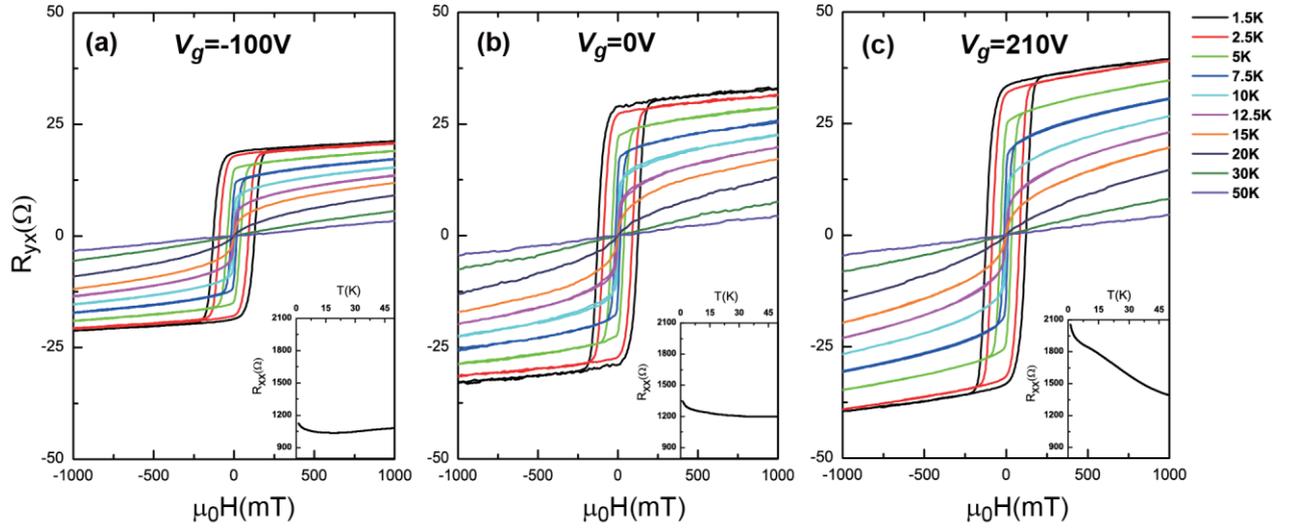

**Fig.4** $\mu_0H$ dependence of the $R_{yx}$ at varied $T$ for **(a)** $V_g$= -100V, **(b)** 0V and **(c)** +210V. Insets of **(a)**, **(b)** and **(c)**, are the respective the $R_{xx}$ vs $T$ curves.

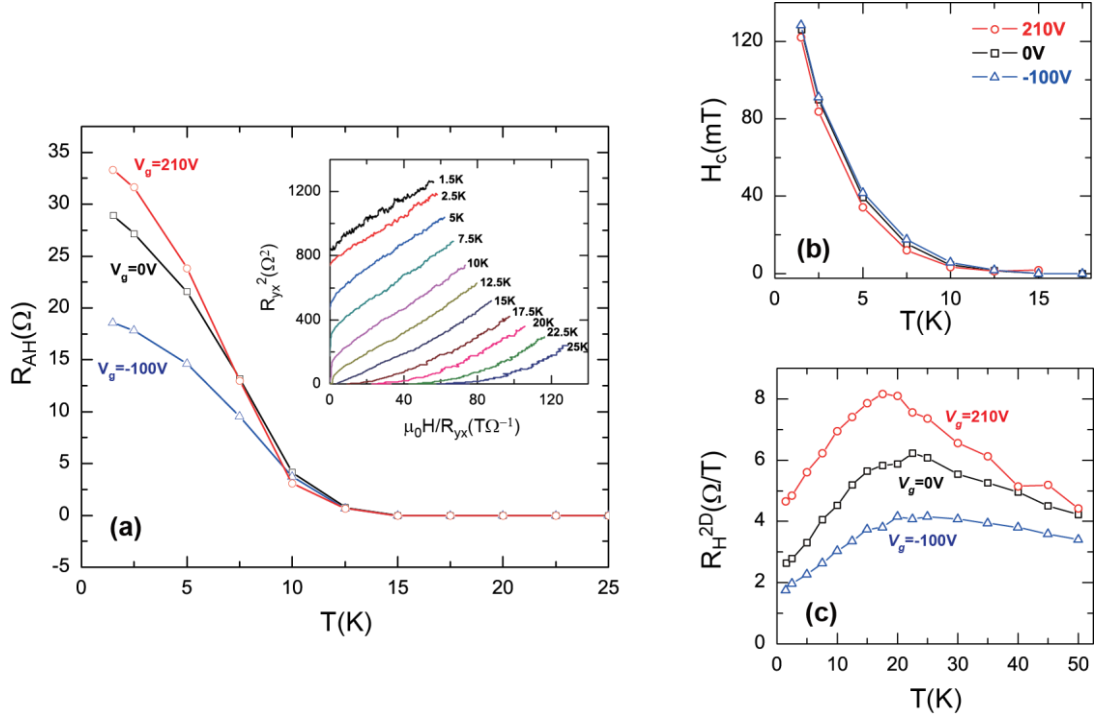

**Fig.5 (a)** *T* dependence of spontaneous Hall resistance ($R_H$) for $V_g$=-100V, 0V and +210V. Inset, Arrot plots for the samples at $V_g$ =0V. **(b)** *T* dependence of coercivity field ($H_c$) for $V_g$=-100V, 0V and +210V. **(c)** *T* dependence of the ordinary Hall coefficient ($R_O^{2D}$) for $V_g$=-100V, 0V and +210V. Data at $V_g$=-100V, 0V and +210V are shown by blue up triangles, black squares and red circles, respectively.